# Extremely Large Nondegenerate Nonlinear Index and Phase-Shift in Epsilon-Near-Zero Materials

Sepehr Benis[1,2], Natalia Munera[1], Sanaz Faryadras[1], Eric W. Van Stryland[1], and David J. Hagan[1,*]

[1]*CREOL, College of Optics and Photonics, University of Central Florida, Orlando, Florida 32816, USA*
[2]*Currently at SiLC Technologies, Inc., Monrovia, CA 91016*
*\*hagan@creol.ucf.edu*

**Abstract:** Epsilon-near-zero (ENZ) materials have emerged as viable platforms for strong nonlinear optical (NLO) interactions. The NLO phase shift in materials exhibiting an ENZ condition is extremely large, however, direct experimental measurements of the magnitude and time dynamics of this phenomenon, particularly nondegenerate NLO phase shifts, have so far been lacking. Here, we directly measure the NLO phase shift of an Indium-Tin-Oxide (ITO) thin film using three different techniques. By characterizing the excitation-induced, time resolved Beam Deflection (BD) of a probe beam, we measure the nondegenerate NLO effects, allowing a separate determination of the effects of excitation and probe wavelengths on the NLO phase shift as they are varied across the ENZ region. These experiments reveal that having the probe pulse centered at ENZ greatly contributes to this enhancement; however, the NLO phase shift is less sensitive to the excitation wavelength, which only slightly enhances the nonlinearity for obliquely incident TM-polarized light. We also find that the spectral shift of the probe pulse induced by the excitation follows both the magnitude and time dynamics of the NLO phase shift measured via the BD experiments. We observe large, ultrafast cross-phase modulation in agreement with a redistribution of carriers in the conduction band. Finally using the Z-Scan method, we measure the degenerate nonlinear refraction at ENZ near normal incidence. The results of all three measurements agree, revealing a gigantic sub-picosecond NLO phase shift in ITO. At its largest, we consistently measure an effective induced index change that is greater than the linear index.

## 1. Introduction

In nonlinear optical (NLO) devices such as all-optical signal processors, a long propagation length is typically required to obtain NLO interactions leading to reasonable NLO phase shifts on the order of π radians [1-3]. This is mainly because the inherent NLO coefficients are relatively small in typical NLO materials such as semiconductors [4, 5], crystals [2, 6], and organic polymers [7-10], which in turn results in NLO interactions that are considered as a perturbation to linear optical properties. To mitigate this, there have been numerous studies optimizing the light-matter interaction to enhance optical nonlinearities such as localized enhancement in optical intensity [11-13] and nonlinear and structured materials with engineered dispersion [14-16]. Efforts have also been dedicated to exploring ENZ materials, at wavelengths where the real part of the permittivity passes through zero, for realizing NLO effects such as large nonlinear refraction (NLR) and absorption (NLA) [17-21], all-optical modulation [21-25], tailored resonance engineering [20-24, 26], adiabatic frequency conversion [18-22], and dynamic quantum physical effects [18, 19, 23, 27, 28].

Transparent conducting oxides (TCO) including tin-doped indium oxide (ITO) and aluminum-doped zinc oxide (AZO) are natural choices for such studies particularly because they exhibit their ENZ condition in the telecommunication window [19, 23, 29]. The ENZ condition in TCOs occurs in proximity to the bulk plasma frequency, $\omega_p/\sqrt{\varepsilon_\infty}$, where the optical behavior is described via the Drude model by $\varepsilon(\omega) = \varepsilon_\infty - \frac{\omega_p^2}{\omega^2 + i\gamma\omega}$, where $\omega$ is the optical angular frequency, $\varepsilon_\infty$ is the high-frequency permittivity, $\omega_p$ is the plasma frequency, and $\gamma$ is the damping factor. These materials can exhibit a notable NLO phase shift, $\Delta\varphi$, in the vicinity of the ENZ region, manifested by the effective NLR for photon energies below the bandgap [17, 19]. Additionally, this extremely large sub-picosecond NLO phase shift may allow obtaining a relatively large spectral shift, $\Delta\lambda$, of the incident light [22, 25, 28, 30]. This simultaneous tunability of both phase and spectrum may be utilized in all-optical signal modulators enabling light-induced switching and frequency synthesizing applications. Additionally, optically-induced spatial and temporal modifications (space-time tunability) can be used to emulate quantum physical effects allowing the observation of such phenomena in optics [14, 22, 27, 28, 30, 31].

NLR is typically characterized via the Kerr nonlinear coefficient, $n_2$, or third-order nonlinear susceptibility, $\chi^{(3)}$ [1, 2, 32]. These descriptions have been shown to provide reasonable insight on the large magnitude of NLR explained via the inverse dependence of the $n_2$ coefficient on the linear index, and thus the enhanced NLR at ENZ [1, 17, 18, 22, 23]. We note, however, that the original theories used for explaining the enhancement strictly only apply to bound-electronic nonlinearities while the dominant NLO effect in TCOs has been experimentally shown to be from redistribution of free-carriers [18, 19, 29, 30, 33-35]. Typically, NLR is extracted from the accumulated NLO phase shift, which is a measurable quantity in NLO experiments. However, single-beam geometries such as Z-Scan are unable to directly measure the time-dynamics of $\Delta\varphi$, and only allow measurement of degenerate NLO coefficients [36, 37]. Additionally, the usual excite-probe experiments cannot separately measure the NLR and NLA, and their time-dynamics are typically determined indirectly by measuring both transmitted and reflected fields at different delays, and thus decoupled to extract the associated coefficients [21, 33]. These techniques also require separate characterization of the substrate to accurately extract the NLO signal of thin films [38, 39]. To overcome these limitations, we employ our Beam-Deflection (BD) technique [28-30, 40, 41] to directly characterize the magnitude, time dynamics, and relative polarization-dependence of the nondegenerate NLO phase shift, $\Delta\varphi$, of a weak probe induced by a strong excitation beam. BD allows us to differentiate refractive mechanisms contributing to $\Delta\varphi$, while characterizing the polarization-dependence of the NLO phase shift at relative temporal delay, $\tau_D$, between the excite and probe pulses. BD can also distinguish between different mechanisms contributing to the NLO response by characterizing the relative polarization-dependent NLO phase shift as well as decoupling them in the temporal domain [28-30, 40, 41]. Thus, it can be used in separately determining the contribution of each mechanism to the NLO phase shift.

In this work, we present an extensive experimental study of NLO effects in ITO around the ENZ condition. Our experimental work is comprised of nondegenerate measurements of the NLO phase shift via BD, degenerate measurements at ENZ via Z-Scans, and cross-phase-modulation (XPM) experiments to characterize the optically-induced spectral shift. Using BD and Z-Scan measurements, we show that a large NLO phase shift can be achieved at ENZ for combinations of excitation and probe beam wavelengths, while XPM experiments independently validate our results with remarkable agreement. Our findings reveal an extremely large effective NLR consistent with a hot-carrier redistribution mechanism from excitation via photon energies below the bandgap [18, 29, 30, 34].

## 2. Experiments
*2.1 Linear optical characterization*

We characterize the permittivity of the ITO thin film (310 nm thickness) via spectroscopic ellipsometry, Fig. 1(a) (experimental details in Supplementary Note 1). The ENZ condition occurs around the zero-crossing of Re($\varepsilon$), defined as $|\text{Re}(\varepsilon)| \leq 1$, shown by the red shaded region in Fig. 1(a). At ENZ ($\lambda_{\text{ENZ}} \approx 1250$ nm), the complex index is $\eta_{\text{ENZ}} \approx 0.39 + i0.39$, exhibiting refractive index values smaller than unity and loss values relatively smaller than that of metals and plasmonic materials [13, 42-44]. We note that a lossless ENZ material exhibits an infinite impedance causing the light to be fully reflected from the surface without any penetration to the material. Thus, a perfect ENZ system cannot transfer energy and information resulting in a vanishing group velocity, $v_g \to 0$. As a result, light only propagates within the ENZ medium while the inherent absorptive loss is nonzero. Moreover, ITO is intrinsically dispersive at ENZ akin to working near a resonance where the group index, described by $n_g = c/v_g = n_0 + \omega dn_0/d\omega$, is significantly larger than the linear refractive index, $n_0$, as shown in Fig. 1(b). The group index further enhances compared to $n_0$ for materials with smaller absorptive loss [45-47].

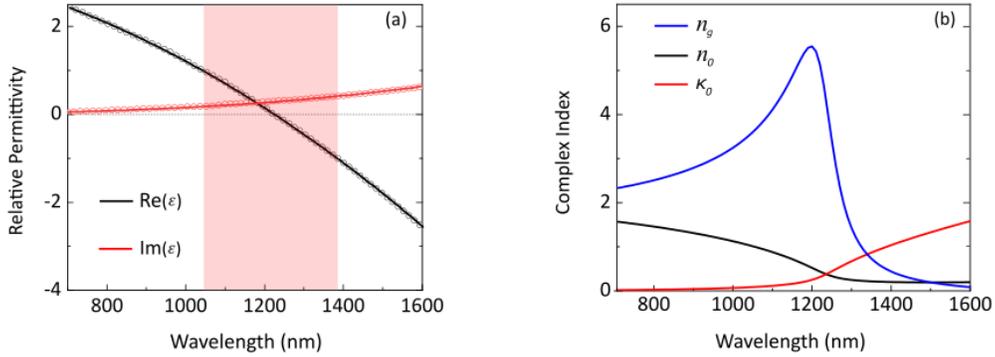

Fig. 1. (a) Real and imaginary parts of the electric permittivity of 310 nm thick ITO (Präzisions Glas & Optik GmbH) measured by ellipsometry and fit by the Drude model ($\varepsilon_\infty = 3.65, \omega_p = 2\pi \times 469 \times 10^{12}\ rad/s$, and $\gamma = 0.041\omega_p$). Red shaded region is where $|\text{Re}(\varepsilon)| \leq 1$, also called the ENZ region. (b) Calculated values of the refractive index (red), extinction coefficient (black), and group index (blue) directly from the measured complex permittivity.

*2.2 Nonlinear optical characterization*

We perform BD measurements for multiple combinations of excitation and probe wavelengths to directly measure the NLO phase shift of the probe beam. For our first set of measurements, we fix the probe wavelength at 775 nm and vary the excitation wavelength across the ENZ, as shown in the shaded region in Fig. 1(a). In this regime, the excitation beam experiences an enhanced absorption mechanism occurring for an obliquely incident TM-polarized light. We perform measurements for two angles of incidence to quantify the effect of enhanced absorption on the redistribution of carriers, and hence, the NLO phase shift. In the second set of measurements, we investigate the nonlinear phase shift and ultrafast spectral shift of the probe while we fix the excitation at 775 nm and tune the probe wavelength across the ENZ region. These two sets of measurements separately characterize the enhancement mechanism for each circumstance, as discussed below, where 1) the enhancement is due to the excitation pulse being at ENZ, which results in a carrier redistribution that only slightly increases the nonlinearity, and 2) the probe pulse is centered around ENZ, which results in a large enhancement of the nonlinear phase shift.

BD becomes useful, particularly because of its ability to temporally resolve optical nonlinearities by directly measuring the nonlinear phase shift for different delay times between the two pulses, and to characterize the NLO induced anisotropy allowing us to determine the physical origin of the NLO effect. This becomes more apparent while distinguishing the signal from the ITO film and substrate by characterizing the induced anisotropy of the NLO response as well as differentiating them in the temporal domain. To fully benefit from the enhanced nonlinearity, we also perform measurements while both excitation and probe beams are in the ENZ region. We find that large and ultrafast $\Delta\varphi$ is achievable for smaller excitation irradiances when both excitation and probe beams are near the ENZ wavelength. For this case, the wavelength shift is also larger compared to the previous cases we measured, particularly because the magnitude of the nonlinear phase shift is significantly enhanced. We also perform degenerate Z-Scan measurements at 1250 nm, which we compare our results to BD measurements performed at slightly nondegenerate wavelengths where the carriers are excited at 1250 nm and the NLO phase shift is probed at 1200 and 1300 nm.

Upon absorption of photons with sub-bandgap energy, the free-carriers will be redistributed in the conduction band resulting in an increase in the kinetic energy of electrons. This effectively modifies the effective mass of electrons by imposing a red shift on the plasma frequency, $\Delta\omega_p$, which, in turn, causes an increase in the permittivity, $\Delta\varepsilon \approx -2(\omega_p/\omega^2)\Delta\omega_p$, and an intensity-dependent change in the effective refractive index, $\Delta n_{\text{eff}}$. This modification is manifested by inducing a phase shift, $\Delta\varphi = k_0 l \Delta n_{\text{eff}}$, on the propagating light described via an effective nonlinear refraction, $\Delta n_{\text{eff}}$, where $l$ is the physical thickness and $k_0$ is the wavevector of light in free-space. Here, we use the description based on the effective change in the refractive index, as defined by $\Delta n_{\text{eff}} = \Delta\varphi/k_0 l$, where $\Delta\varphi$ is the experimentally measured NLO phase shift. We will also discuss a theoretical perspective later based on the influence of dispersion of ITO, explained via the group velocity and phase velocity, on the enhanced sensitivity of the nonlinear phase shift to the induced changes in material permittivity. Additionally, the time-dependent nature of the nonlinear phase-shift results in a spectral shift, $\Delta\lambda$, where its magnitude can be determined based on the strength and temporal dynamics of $\Delta\varphi$ as it can be fully described by means of XPM phenomena.

## 3. Measurements
### 3.1 Role of excitation wavelength on NLO phase shift

Redistribution of hot carriers in the conduction band and subsequently the modification of the effective refractive index, $\Delta n_{\text{eff}}$ (as previously defined), and thus phase shift, $\Delta\varphi$, is directly related to the strength of absorption of light by the free-carriers. At ENZ, the absorption is enhanced due to the small value of the linear refractive index, $n_0$, and large group index, $n_g$, resulting in slow light propagation, as discussed in [34, 48-50]. Inherently, this means that the excitation pulse interacts with the material for a longer time, hence a more efficient light-matter interaction occurs within the material thickness [34]. We also note that absorption is further enhanced for an obliquely incident TM-polarized excitation light. This behavior, however, is due to the more efficient coupling of the normal component of the electric field to the longitudinal plasma oscillations within the film, analogous to the Berreman effect for the phonon polariton mode at the air/dielectric interface [51-54].

To measure the dependence of the magnitude of the NLO phase shift on the extent of carrier redistribution, we perform a set of nondegenerate measurements utilizing our BD method to directly measure the NLO phase shift. In Fig. 2, we show BD measurements performed on ITO at normal incidence for the excitation wavelength, $\lambda_e$, centered at $\lambda_e = 1250$ nm and peak incident irradiance of $I_e = 57 \text{ GW/cm}^2$ while the weak probe is fixed far away from the ENZ region at $\lambda_p = 775$ nm. We choose this wavelength for the probe to avoid any enhancement related to the ENZ condition and only measure the NLO phase shift associated with the efficient carrier redistribution mechanism due to excitation at ENZ. We keep the incident excitation irradiance below the saturation threshold of the NLO response, which is observed at for $I_e > \sim 75$ GW/cm$^2$ (Supplementary Note 3). We perform similar measurements on the glass substrate to scale the measurements of the nonlinear phase shift of ITO, $\Delta\varphi_{\text{ITO}}$, to a known reference, in this case $\Delta\varphi_{\text{SiO}_2}$, to calibrate our experimental apparatus. We also employ BD to accurately characterize the polarization anisotropy of the NLO mechanisms by separately measuring the NLO phase shift for parallel and perpendicular relative polarizations of excitation and probe pulses. We find that unlike third-order bound-electronic nonlinear mechanisms exhibiting anisotropy [40, 55, 56], $\Delta\varphi_{\text{SiO}_2,\parallel} = 3\Delta\varphi_{\text{SiO}_2\perp}$, the signal from ITO is equal for both polarizations at longer delays, as shown in Fig. 2(a), where there is no contribution from the substrate to the nonlinear phase-shift. This indicates that the dominant NLO mechanism in ITO is from non-instantaneous hot carrier redistribution as opposed to that in SiO$_2$ where it is from the instantaneous bound-electronic nonlinearity.

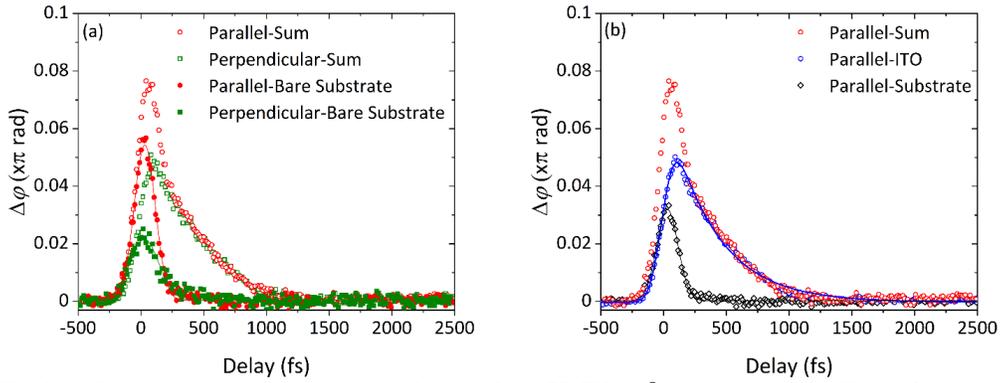

Fig. 2. (a) BD measurements for excitation at 1250 nm ($I_{0,e} \approx 57\ GW/cm^2$) and weak probe at 775 nm for parallel (red) and perpendicular (green) relative polarizations performed on ITO sample (ITO and substrate together – open symbols) and individually on the substrate (closed symbols). Solid lines are fit to the measurements performed on the substrate. (b) BD measurements on ITO sample at parallel polarization (red – same dataset as in (a)) and contribution of the substrate (black – difference from (a) is explained in text) and pure signal for ITO thin film (blue) with parallel polarization extracted from measurements. Solid lines are fit to the ITO as discussed in the text. The extracted signal for ITO for parallel (blue) and perpendicular (not shown here) polarizations are identical within experimental errors indicating no dependence on relative polarization.

The polarization dependence of the NLO phase shift allows us to accurately determine the pure contribution of ITO to the BD signal. Specifically, we first eliminate the signal from ITO by subtracting the parallel and perpendicular total measured signals from the ITO film and the substrate, $\Delta\varphi_{\text{sum},\parallel} = \Delta\varphi_{\text{ITO},\parallel} + \Delta\varphi_{\text{SiO}_2,\parallel}$ and $\Delta\varphi_{\text{sum},\perp} = \Delta\varphi_{\text{ITO},\perp} + \Delta\varphi_{\text{SiO}_2\perp}$, and then we obtain a differential signal, $\Delta\varphi_{\text{sum},\parallel} - \Delta\varphi_{\text{sum},\perp} = \Delta = 2/3\Delta\varphi_{\text{SiO}_2,\parallel}$, purely from the substrate. Since the response from SiO$_2$ is dominated by the instantaneous bound-electronic nonlinearity, the differential signal follows the cross-correlation between excitation and probe, and establishes the zero-delay time, $\tau_D = 0$, between respective pulses. By multiplying $\Delta$ by a factor of 3/2 (black data in Fig 2(b)), and subtracting it from either $\Delta\varphi_{\text{sum},\parallel}$ or $\Delta\varphi_{\text{sum},\perp}$, we can accurately extract the signal due solely from the ITO film, as shown in Fig. 2(b) – blue data. This methodology is particularly useful when the linear absorption and/or nonlinear transmission significantly alter the irradiance of the excitation beam reaching the back surface of ITO while the film is on the front. Our measurements also reveal that the maximum achievable NLO phase shift occurs at a delay, $\tau_{D,\max}$, which is larger than zero. We fit the non-instantaneous response of ITO, shown in Fig. 2(b), as having a pulsewidth-dependent rise-time determined by the temporal integral of a Gaussian excitation pulse with a decay-time, $\tau_f$. The rise is determined by the cumulative nature of the hot carrier redistribution in ITO, while the fall time corresponds to the relaxation mechanism of the redistributed carriers, which is dominated by electron-electron scattering and exhibits a sub-picosecond relaxation dynamic. The nonlinear phase shift in ITO is of the same order as that in SiO$_2$ for parallel polarization

(Fig 2(b) black and blue data); however, the ITO film is ~ 3500 × thinner than the substrate. Therefore, a comparable NLO phase shift is achievable in ITO at orders of magnitude shorter propagation lengths than those in SiO$_2$. The magnitude of the measured NLO phase shift, $\Delta\varphi_{\text{ITO}} = (0.048 + 0.01)\pi$ rad, corresponds to $\Delta n_{\text{eff}} = 0.06 \pm 0.01$ for the given ITO thickness ($l \approx 310$ nm). We note that $\Delta n_{\text{eff}}$ is much smaller than the initial index of refraction, $n_0 \approx 1.52$, at $\lambda_p = 775$ nm calculated from ellipsometry measurements. However, this change corresponds to a ~ 4% change in the magnitude of the linear index, which is considerably larger than that of conventional bulk NLO materials such as glass [57, 58] and semiconductors [4, 56] and is achievable in a subwavelength propagation distance.

To quantify the NLO response of ITO for excitation wavelengths across the ENZ region, we calculate the spectral dependence of the normalized absorbed energy, $A(\lambda, \theta) = 1 - R(\lambda, \theta) - T(\lambda, \theta)$, where $R$ and $T$ are the reflected and transmitted energy calculated via the transfer matrix method, and $\lambda$ and $\theta$ are the excitation beam wavelength and angle of incidence, respectively. Fig. 3(a) represents the absorption versus $\theta$ and $\lambda$ for incident light at TM polarization. We find that absorption enhances at oblique incidence (20° ≤ $\theta$ ≤ 60°) exhibiting a 'Berreman-type' resonant behavior that can be attributed to the enhanced coupling of the normal electric field to longitudinal plasma oscillations [46, 47, 51, 53, 54]. Owing to the larger absorption at ENZ and oblique incidence, the redistribution of carriers is more efficient, and the probe pulse experiences a larger NLO phase shift (see Fig. 3(b)). We note, however, that this enhanced absorption is exclusively dependent on the absorptive nature of the material response and hence can be further enhanced for an engineered material with a larger loss but spanning a broader spectral bandwidth, as predicted in [52, 59]. The plasma oscillations caused by the incident light are inherently related to the singularity of the energy loss of charged particles in the vicinity of the plasma frequency [46, 60]. We note that the imaginary part of the complex index, as shown in Fig 1(b), may suggest that the absorbed energy monotonically increases as the excitation wavelength is increased across the ENZ region; however, the reflection also increases significantly at longer wavelengths due to the increase in loss, albeit the refractive index maintains its small value. This results in maximum absorption close to ENZ as shown in Fig. 3(a). The spectral location of the maximum absorption depends upon the loss value and thickness, and it may be slightly blue shifted from the ENZ wavelength. Consequently, this results in observing the maximum NLO phase shift at a slightly shorter wavelength than ENZ.

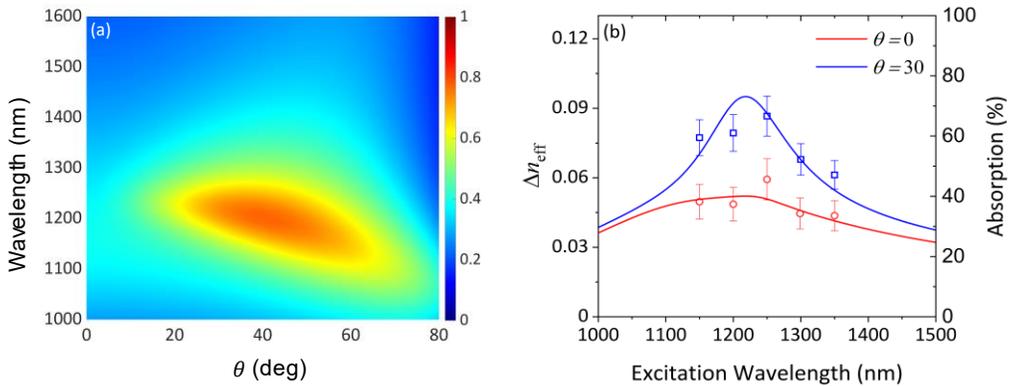

Fig. 3. (a) Normalized absorbed energy $A(\lambda, \theta)$ (right axis - colorbar) calculated via transfer matrix method for TM-polarized light by using complex permittivity of ITO extracted from ellipsometry measurements (solid lines). (b) Extracted $\Delta n_{\text{eff}}$ (left axis) at $\tau_{D,max}$ from nondegenerate BD measurements at each excitation wavelength across ENZ region at a fixed irradiance $I_e = 57 \pm 5 \, GW/cm^2$. The absorption spectrum (right axis) is overlayed on the measured data at the relevant angle of incidence to compare the scaling of the spectral dependence of $\Delta n_{\text{eff}}$ vs absorption spectra.

We perform BD measurements at normal incidence ($\theta = 0°$) and oblique incidence ($\theta = 30°$) for both excitation and probe beams (Supplementary Note 3). We directly calculate $\Delta n_{\text{eff}}$ at each excitation wavelength at an incident irradiance of $I_e = 57 \pm 5$ GW/cm² as summarized in Fig. 3(b). The optically-induced $\Delta n_{\text{eff}}$ reaches the value of $0.09 \pm 0.01$ at ENZ excitation at 30° exhibiting ~ 50% enhancement in the achievable $\Delta n_{\text{eff}}$ compared to that at normal incidence. We note, however, that for thinner ITO films, where absorption is smaller at normal incidence, the enhancement factor may be higher for obliquely incident excitation beam, as previously shown in [59, 61]. We compare the spectral dependence of $\Delta n_{\text{eff}}$ with that of the absorbed energy at each wavelength and incidence angle and see a strong correlation in relative magnitude and shape as shown in Fig 3(b). Our results corroborate that the elevated redistribution of carriers at oblique incidence directly contributes to the enhanced magnitude of the nonlinear phase shift and subsequently $\Delta n_{\text{eff}}$.

## 3.2 Role of probe wavelength on NLO phase shift

Here we directly measure the nondegenerate $\Delta\varphi_{\text{ITO}}$ at normal incidence for probe wavelengths spanning the ENZ region (900 – 1550 nm), while the excitation wavelength is relatively far from ENZ, fixed at $\lambda_e = 775$ nm. We perform these measurements to investigate the NLO phase shift while the probe experiences the ENZ condition. For each probe wavelength, we also perform BD measurements for parallel and perpendicular relative polarizations on the substrate. This allows us to effectively extract the nonlinear phase shift of the ITO film by eliminating substrate contributions with the procedure outlined in section 3.1. Measurements at multiple irradiances for different probe wavelengths are shown in Supplementary Note 4. The irradiance is kept sufficiently low to avoid saturation of the nonlinear response. We note, however, that the saturation occurs at a much higher irradiance compared to when excitation is at ENZ. This is due to the greatly reduced absorption of ITO at $\lambda_e = 775$ nm, $\alpha \approx 4500\ cm^{-1}$, indicating that a higher excitation irradiance is needed to achieve a comparable phase shift and plasma frequency shift within the same thickness. In this case, we observe an extremely large NLO phase shift of the probe pulse at 1250 nm for a peak excitation irradiance of $I_{0,e} = 521$ GW/cm² . We obtain $\Delta\varphi_{\text{ITO}} = (0.21 \pm 0.02)\pi$ rad , corresponding to $\Delta n_{\text{eff}} = 0.42 \pm 0.04$ at $\tau_D \sim 100$ fs , where the probe experiences its maximum phase shift. The magnitude of the maximum $\Delta\varphi_{\text{ITO}}$ is smaller at probe wavelengths of 950 nm and 1550 nm by factors of 4 and 4.4, respectively. Remarkably, the observed effective index change, $\Delta n_{\text{eff}} = \Delta\varphi_{\text{ITO}}/k_0 l$, can reach values close to unity allowing for ~100% modulation of the initial refractive index, $n_o \sim 0.4$. This is considerably greater than the case where the probe was kept far away from ENZ, albeit having a less efficient excitation at $\lambda_e = 775$ nm. This enhanced NLO effect for the fixed excitation absorption is indicative of the pronounced sensitivity of $\Delta\varphi$ to the optically-induced change in plasma frequency.

We simultaneously perform XPM experiments at different temporal delays to separately characterize the optically-induced wavelength shift of the probe, $\Delta\lambda_p$ . Our measurements reveal substantial spectral shifts due to the extremely large magnitude of nonlinear phase shift together with its sub-picosecond relaxation dynamics. For these measurements, we use a fiber-coupled optical spectrum analyzer to collect the probe spectrum transmitted through the film. We perform these measurements for different temporal delays between excitation and probe pulses and observe a red shift while the optical phase increases ($\partial\Delta\varphi/\partial t > 0$) and a blue shift while the optical phase decreases ($\partial\Delta\varphi/\partial t < 0$), as illustrated in Fig. 4. We also compare our XPM measurements with the calculated spectral shift from the time derivative of $\Delta\varphi_{\text{ITO}}$ obtained directly from BD measurements, as shown in Fig. 4(b). We obtain excellent agreement at each individual probe wavelength, as shown in Supplementary Note 3. This agreement between two techniques is satisfying, particularly when there are no fitting parameters used in our approach. We use $\Delta\lambda_p = -2\pi c/\lambda_p^2 \cdot \partial\Delta\varphi_{\text{ITO}}/\partial t$, where $c$ is the speed of light in free-space and calculate $\Delta\lambda_p$ directly from our BD measurements at various excitation irradiances. The data nicely follows the derivative of the phase shift, and we find a red shift maximum of $\Delta\lambda_p \approx 2.6$ nm ($\Delta f_p \approx -0.5$ THz) at $\tau_D = -100$ fs, with a somewhat smaller second blue shift maximum of $\Delta\lambda_p \approx -1.2$ nm ($\Delta f_p \approx 0.23$ THz), at $\tau_D = 380$ fs with $I_e = 521$ GW/cm$^2$, for $\lambda_p = 1250$ nm. We note, however, that the NLO effect can become significantly larger for excitation and probe pulses propagating at the same group velocity [22, 62]. Additionally, the spectral shift may be further enhanced for shorter pulses, which allows obtaining a faster cumulative rise for $\Delta\varphi$.

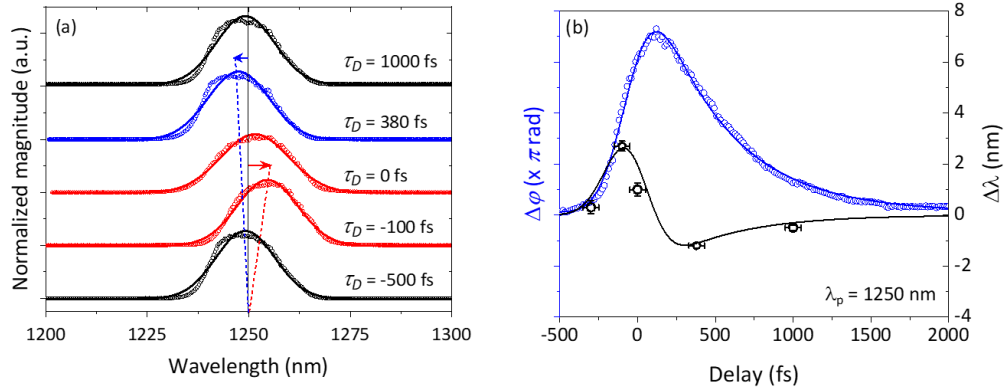

Fig. 4. (a) XPM measurements at different delays between excitation, $\lambda_e = 775\ nm$, and probe, $\lambda_p = 1250\ nm$, at $I_e = 521$ GW/cm$^2$. Red and blue data and fit represent redshift and blueshift, respectively. (b) BD measurements and fit (left axis - blue) for $\lambda_p = 1250\ nm$ at $I_e = 521$ GW/cm$^2$ and (right axis - black) calculation of $\Delta\lambda_p$ based on $\partial\Delta\varphi_{\text{ITO}}/\partial t$, and spectral shift data obtained from (a).

We extract $\Delta n_{\text{eff}}$ directly from the experimentally measured values of $\Delta\varphi_{\text{ITO}}$ and compare them to the calculated spectral dependence of $\Delta n_{\text{eff,calc.}} = \sqrt{\varepsilon + \Delta\varepsilon} - \sqrt{\varepsilon}$, where $\Delta\varepsilon \approx -2(\omega_p/\omega^2)\Delta\omega_p$. We note that $\Delta\omega_p$ is the induced shift in plasma frequency and not to be confused with the shift in probe frequency. We also fit the scaling of the spectral dependence of $\Delta n_{\text{eff,calc.}}$ to our measurements and determine $\Delta\omega_p$ at each excitation irradiance. The calculated spectral dependence of the effective NLR, $\Delta n_{\text{eff,calc.}}$, exhibits remarkable agreement with our experimental results, as shown in Fig. 4(a). From the fit, we obtain $\Delta\omega_p/\omega_p$ of 3.1%, 5.8%, and 9% relative to the initial value of the plasma frequency at each excitation irradiance ($I_e = 183, 337,$ and $521$ GW/cm$^2$), respectively. These values show the expected linear dependence of the shift with irradiance. We note that calculated $\Delta\omega_p$ values are relatively small, albeit we measure extremely large $\Delta\varphi_{\text{ITO}}$, and subsequently $\Delta n_{\text{eff}}$. We also calculate the time derivative of $\Delta\varphi_{\text{ITO}}$ to characterize the spectral dependence of the red shift and blue shift of the probe (Details in Supplementary Note 4). The maximum magnitude of the achievable redshift at ENZ is around 5 times larger than that at 950 nm and 10 times larger than at 1550 nm corroborating that the spectral shift is also enhanced with the probe centered at ENZ.

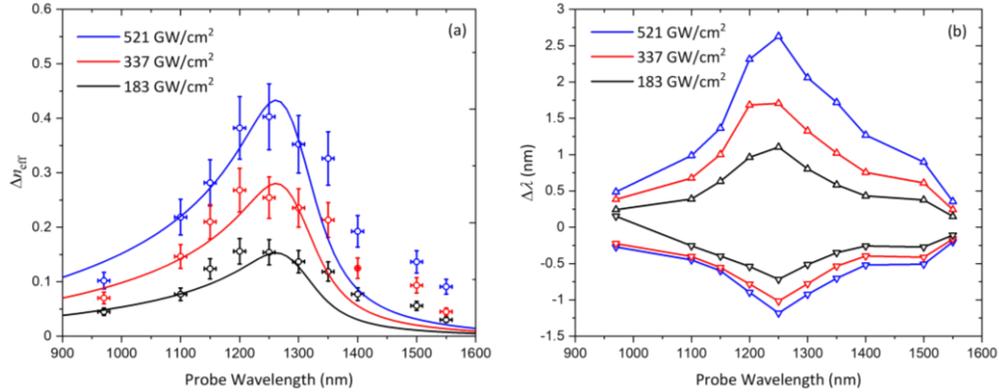

Fig. 5. (a) Comparison of the extracted effective NLR, $\Delta n_{\text{eff}}$, with the predicted spectral dependence calculated from the Drude model for an excitation at 775 nm, and (b) spectral dependence of the optically induced red shift (up triangles) and blue shift (down triangles) of the probe pulse directly calculated from $\partial \Delta \varphi_{ITO}/\partial t$ measured by BD.

We note, however, that the linear absorption of the excitation beam at 775nm ($\alpha \approx 4500\ cm^{-1}$) is very small, thus negligible excitation depletion occurs while it propagates within the ITO film. However, in this case, the efficacy of carrier redistribution is smaller, requiring higher irradiances to achieve a similar performance to excitation at ENZ. The carrier excitation becomes further constrained due to the injection of excess carriers from valence to conduction band via two-photon absorption (2PA, $\alpha_2$), where $\alpha_2 = (3.8 \pm 0.5)$ cm/GW at $\lambda_e = 775$ nm as measured by Z-Scan and presented in Supplementary Note 6 and discussed in section 3.4. This effectively increases the number of free carriers in the conduction band and negatively contributes to the NLO phase shift of the probe as the excitation beam irradiance increases. The decrease of the NLO phase shift via 2PA is two-fold, firstly, from the reduction of excitation irradiance within the film, and secondly, due to the excess cold free-carriers excited to the conduction band. Thus, the minor disagreement between calculated $\Delta n_{\text{eff}}$ from the Drude model and our measurements, shown in Fig. 5(a), may be attributed to the contribution of excited bound carriers to the free-carrier density originated by the 2PA mechanism.

To further enhance the nonlinear phase shift, we perform measurements with the excitation fixed at $\lambda_e = 1250$ nm and the probe varied across the ENZ region. We find that at $\lambda_p = 1200$ nm, the probe experiences an extremely large nonlinear phase shift, $\Delta \varphi_{\text{ITO}} = (0.44 \pm 0.04)\pi$ rad, corresponding to $\Delta n_{\text{eff}} = 0.9 \pm 0.08$ at $I_e = 100$ GW/cm$^2$. We note that the extremely large nonlinear phase shift and near-unity $\Delta n_{\text{eff}}$ is achievable at a much smaller irradiance than obtained with excitation at $\lambda_e = 775$ nm, i.e., a ~10 × better efficiency of the carrier redistribution mechanism for excitation at ENZ due to larger absorption, $\alpha \approx 39200\ cm^{-1}$, as also predicted in [28]. Fig. 6(a) compares the spectral dependence of the measured effective NLR to the calculated $\Delta n_{\text{eff}}$ values from the Drude model. Similar to our previous analysis, by fitting the calculated spectral dependence for an arbitrary shift in plasma frequency to the measured $\Delta n_{\text{eff}}$ we find relative values of $\Delta \omega_p/\omega_p$ of 3.1%, 6.1%, 9.2%, 12.4%, and 18.5% for each excitation irradiance ($I_e = 17, 33, 50, 67$, and $100$ GW/cm$^2$), respectively showing a linear dependence on irradiance. This validates that $\Delta \omega_p$ is larger for excitation and probe at ENZ compared to previous cases where one pulse was kept away ENZ.

We also perform XPM experiments and obtain excellent agreement between our values and the directly extracted spectral shift values from our BD measurements. We find that the spectral shift is also enhanced in this case ($\lambda_e = 1250$ nm, $\lambda_p = 1200$ nm), yielding an 8.1 nm ($\Delta f_p \sim -1.7$ THz) red shift and 3.8 nm ($\Delta f_p \sim 0.8$ THz) blue shift at $I_e = 100$ GW/cm$^2$. We, therefore, obtain ~16 × better performance compared to when $\lambda_e = 775$ nm, giving larger frequency shifts at more reasonable excitation irradiances. We note that both the magnitude and temporal dependence of $\Delta\varphi_{\text{ITO}}$ contribute to the magnitude of the spectral shift, as described earlier via $\Delta\lambda_p = -2\pi c/\lambda_p^2 \cdot \partial\Delta\varphi_{\text{ITO}}/\partial t$. This dependence of the spectral shift on the temporal evolution of $\Delta\varphi$ is entirely due to the XPM phenomena. We note that the frequency shift is extremely fast, although non-instantaneous, but occurring on a sub-picosecond time scale, which may allow frequency conversion with very high modulation speeds.

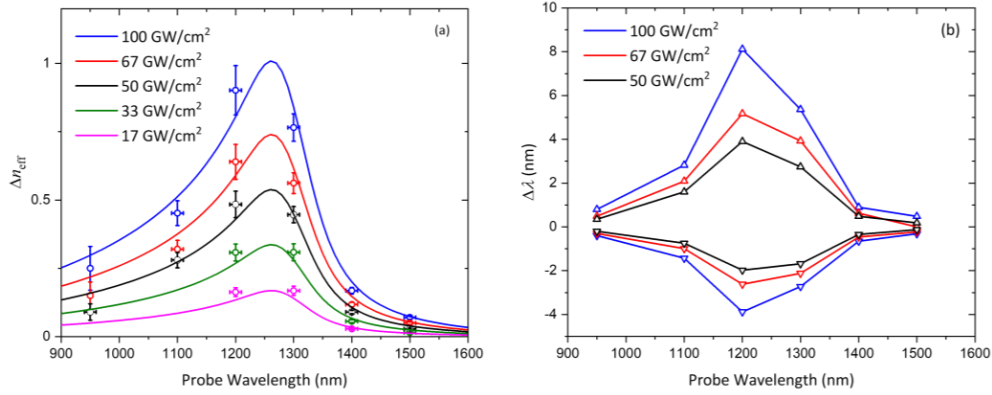

Fig. 6. (a) Comparison of the extracted effective NLR, $\Delta n_{\text{eff}}$, from BD measurements with the predicted spectral dependence calculated from the Drude model for an excitation fixed at 1250 nm, and (c) spectral dependence of the optically-induced red shift (up triangles) and blue shift (down triangles) of the probe pulse as directly calculated from $\partial\Delta\varphi_{\text{ITO}}/\partial t$, where $\Delta\varphi_{\text{ITO}}$ is measured via BD (analogous to Fig. 5(b)).

3.4 Z-Scan

We note that our measurements with both ENZ excitation and probe are intentionally kept slightly nondegenerate to minimize the artifacts present in degenerate two-beam experiments [63]. We, therefore, perform Z-Scan measurements at 1250 nm (Supplementary Note 6) to compare the results of a single-beam degenerate experiment to our BD measurements at $\lambda_p = 1200$ nm and 1300 nm [36, 37]. For this comparison, we deliberately choose the $\Delta n_{\text{eff}}$ values obtained via BD at zero-delay, $\tau_D = 0$, where the two pulses are entirely overlapping to mimic the interaction geometry of the single-beam Z-Scan technique. The agreement between three techniques (BD, XPM and Z-Scan) is extraordinary, particularly when there are no fitting parameters used in our approach.

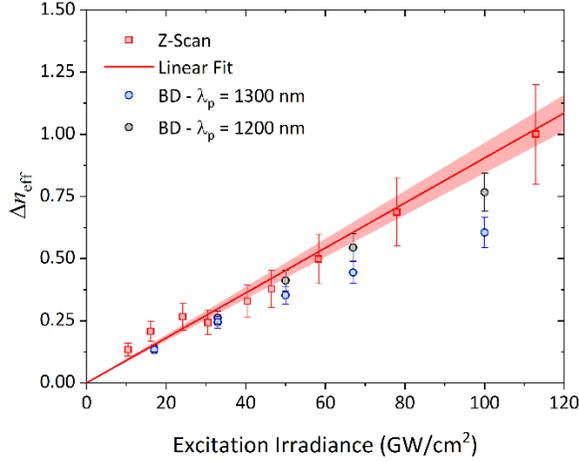

Fig. 7. Comparison of the measured $\Delta n_{\text{eff}}$ via degenerate Z-Scan at ENZ condition, $\lambda = 1250\ nm$, with values obtained at zero delay, $\tau_D = 0$, from the BD measurements at probe wavelengths, $\lambda_p = 1200\ nm$ and $\lambda_p = 1300\ nm$ for different excitation, $\lambda_e = 1250\ nm$, irradiances. The $\Delta n_{\text{eff}}$ extracted from BD is slightly lower than that of Z-Scan for degenerate NLR. The red shaded region exhibits the 90% confidence level of the linear fit to Z-Scan measurements (red squares).

## 4. Discussion

As discussed earlier, the optically-induced effective NLR, $\Delta n_{\text{eff}}$, is typically extracted from the experimentally measured NLO phase shift, $\Delta n_{\text{eff}} = \Delta\varphi/k_0 l$. We use this notation in analogy with optical waveguides, where $\Delta n_{\text{eff}}$ is a measure of the change in the modal index, in which dispersion information of the guided mode is needed to relate $\Delta n_{\text{eff}}$ to the NLR of the guiding material, $\Delta n$. In those cases $\Delta n_{\text{eff}}$ gives the index change pertinent to the propagating mode as opposed to the change in the intrinsic material refractive index, $\Delta n$ [14, 18, 49, 50, 62]. The ENZ condition in TCOs also appears at spectral regions where the material exhibits a highly dispersive behavior. In this case, however, $\Delta n$ and $\Delta n_{\text{eff}}$ are not experimentally distinguishable. $\Delta n_{\text{eff}}$ is indeed the practical measure of the nonlinear phase shift, which is the essential quantity in designing NLO devices.

The large dispersion around ENZ fundamentally influences the effective interaction time due to the apparent contrast between the phase velocity (refractive index less than unity) and slowed-down group velocity (large group index), as shown in Fig. 1(b). In this case, the energy transfer between excitation light and free-carriers significantly increases; while simultaneously the NLO phase shift of the ENZ probe also becomes extremely sensitive to changes in permittivity [18, 35, 46, 49, 50, 62]. To understand this effect further, we establish a formalism that relates $\Delta\varphi$ to the dispersion, *i.e.* phase and group velocities, and the effective propagation length, $l_{\text{eff}}$. This effective length includes factors relevant to the material dispersion and is defined as $\Delta\varphi = k_0 l_{\text{eff}} \Delta n$, where $\Delta n$ may be used to describe the "material" index change. We note that ENZ materials are intriguing mainly due to their ability to exhibit relatively small group velocity, $v_g = c/n_g$, at frequencies where the phase velocity, $v_p = c/n_0$, is larger than usual. This condition dramatically increase the phase sensitivity caused by any given change in permittivity, as also previously shown in TCOs and photonic structures [18, 35, 46, 49, 50, 62].

To better understand the increase in phase sensitivity, we illustrate in Fig. 8 two dispersion curves in $f - k$ space for an ITO thin film with no NLO effects (blue), and for an induced relative change of plasma frequency of $\Delta\omega_p/\omega_p = 18.5\%$ (red). The initial (point A) and final (point B) pairs of $(f, k)$ are presented in accordance with our slightly nondegenerate BD and XPM measurements discussed in section 3.3. There we have $\lambda_p = 1200$ nm and $\lambda_e =$

1250 nm for an incident excitation irradiance of $I_e = 100$ GW/cm$^2$, yielding $\Delta f_p \approx -1.7$ THz and $\Delta k_p = (4.45 \pm 0.4)~\mu m^{-1}$ ($\Delta\varphi_{\text{ITO}} = (0.44 \pm 0.04)\pi$ rad). We note that in a typical NLO interaction, the material is perturbed both in time and space, and thus neither frequency nor wavevector are conserved. However, we may ignore the relatively small vertical transitions in the frequency domain and only consider horizontal transitions between dispersion curves (dotted line in Fig. 8). Note that we also keep the pulsewidth in our experiments long enough (i.e., above 150 fs to have a relatively narrow bandwidth) to avoid any confounding effects from the group velocity dispersion of ITO.

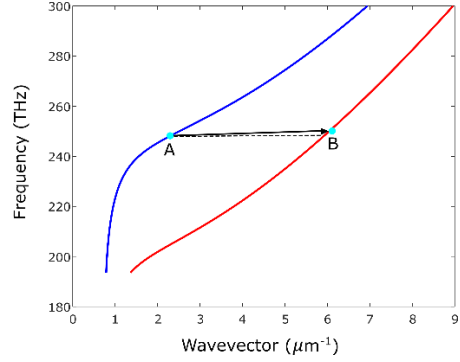

Fig. 8. Dispersion of ITO thin film with no NLO effects (blue) calculated via the Drude model with parameters extracted from ellipsometry measurements. The red curve exhibits the dispersion with a nonlinearly-induced shift in the plasma frequency, $\Delta\omega_p/\omega_p = 18.5\%$. The black solid arrow represents the transition from the initial state of material in the $f - k$ plane (point A) to the final state (point B) where both frequency and wavevector are modified. The black dotted line is the simplified transition from A to B while ignoring the shift in frequency.

We calculate the scaling of the interaction length under a Drude dispersion described by $\omega(k) = ck/n$. More precisely, for an induced change in the dispersion of the material, the NLO phase shift may be written as

$$\Delta\varphi_{\text{ITO}} = \Delta l \approx \frac{\partial k}{\partial f}\Delta f l, \tag{1}$$

where $\Delta k$ is the magnitude of shift in wavevector and $\Delta f$ is the optically-induced change in frequency for a given induced shift in plasma frequency, $\Delta\omega_p$. We note that this approximation is only valid while considering a minimal change in frequency, as evidenced by our experimental results and explained earlier where we observed $\Delta f \ll f$. The shift in frequency can be further approximated as $\Delta f/f \approx -\Delta n/n$ from perturbation analysis [64], which is also sometimes referred to as the time-refraction phenomenon [64, 65]. We note that, in this approximation, $\Delta n$ is a measure of NLR of the ITO thin film independent of the linear dispersion of ITO, which is now embedded into $l_{\text{eff}}$. Using the definition of group velocity, $v_g = 2\pi\partial f/\partial k$, and substituting the above approximation in Eq. (1), we obtain

$$\Delta\varphi_{\text{ITO}} \approx -2\pi\frac{l}{v_g}f\frac{\Delta n}{n}, \tag{2}$$

where $v_g = c/n_g$ is the group velocity of light. We can also rewrite this equation in terms of both group velocity and phase velocity, $v_p = c/n_0$, as $\Delta\varphi \approx -(v_p/v_g)lk_0\Delta n = -(n_g/n_0)lk_0\Delta n$. We use the ratio of $S(\lambda_p) = v_p/v_g = n_g/n_p$ at the light wavelength to describe the strength of group velocity reduction, *i.e.* the slow-down factor. The slow-down factor is typically very small (near unity) for conventional NLO materials, where the spectrum of interest is far away from material resonances.

This factor can be enhanced further in ITO films for either ENZ probe or excitation pulses, where each contribute to the NLO enhancement mechanism differently. For an ENZ probe, the effective NLR of ITO is larger as evidenced by our experiments. This intrinsic enhancement may be described by the extended effective interaction length while considering that now the calculated change in the material index may not be so large. In other words, the improvement of light-matter interaction described by the extended length, $l_{\text{eff}}(f) = lS(f)$, may be used as an alternative interpretation of the resonant behavior in the NLO phase shift. This enhancement manifests its effect as a higher sensitivity of NLO phase shift to the induced changes in plasma frequency.

Additionally, the ENZ condition influences the propagation of the excitation pulse by increasing the interaction time and thus improving the carrier redistribution [19, 20, 22, 30, 34, 46, 48, 59]. Thus, the amount of absorbed energy also scales with the slow-down factor calculated at the excitation wavelength, hence an improved NLO effect. This behavior can be understood by calculating the difference between the injected energy density and the density after propagating through the sample as fully explained in [34]. We note that the enhancement mechanisms of NLO effects in TCOs are fundamentally different from those occurring in structured materials or organic molecular compounds, which are resorting to enhanced NLO phenomena such as resonant third-order nonlinearities and local field enhancement, respectively. In ENZ materials, the two scaling factors at probe and excitation wavelengths result in a longer propagation length and subsequently a large NLO phase shift. We note that, besides this enhancement mechanism, the induced change in effective index, $\Delta n_{\text{eff}}$, is still very large in TCOs enabling one to achieve near unity values in the effective nonlinear refractive index. This means that $\Delta n_{\text{eff}}$, including all possible enhancement mechanisms, can be experimentally determined to be near unity as shown in our BD measurements.

## 5. Conclusions

In conclusion, we have experimentally characterized the degenerate and nondegenerate NLO phase shift of ITO for multiple spectral combinations using three different techniques. Our temporal and polarization-dependent BD measurements demonstrate that carrier-induced nonlinearities are dominant in the NLO response. We observe extremely large effective index changes, $\Delta n_{\text{eff}}$, on the order of unity near the ENZ condition and we note that even outside the ENZ region the magnitude of $\Delta n_{\text{eff}}$ is much larger than that of other highly nonlinear materials such as third-order nonlinear effects in AlGaAs [66, 67] and chalcogenide glass [57]. Additionally, we simultaneously characterized the effect of optically-induced spatial and temporal modulation of the material permittivity, by simultaneously measuring $\Delta \varphi$ and $\Delta \lambda$ by BD and XPM experiments. We obtained excellent agreement between the three different experimental techniques. We also presented a theoretical perspective on the significance of group velocity and phase velocity on the superior sensitivity of the NLO phase shift to the induced changes in material index. This explanation unveils the role of slow-light propagation in the NLO response at ENZ. The outcomes of this work can be used for efficient dispersion engineering close to material resonances and devising photonic systems towards improving light-matter interaction.

**Funding.** Air Force Office of Scientific Research (MURI: FA9550-20-1-0322, MURI: FA9550-21-1-0202); Army Research Lab (W911NF-15-2-0090)

**Acknowledgments.** We gratefully thank Nathaniel Kinsey (Virginia Commonwealth University) for the fruitful conversations and his valuable comments on this work.

**Disclosures.** The authors declare no conflicts of interest.

**Data availability.** Data underlying the results presented in this paper are all presented within this article.

**Supplemental document.** See Supplement 1 for supporting content.

# Extremely Large Nondegenerate Nonlinear Index and Phase-Shift in Epsilon-Near-Zero Materials: supplemental document

**Supplementary Note 1: Characterization of linear optical properties of ITO.**

We performed spectroscopic ellipsometry on a commercially bought thin layer of ITO (thickness 310 nm - Präzisions Glas & Optik GmbH) deposited on a glass substrate (thickness 1.1 mm) to characterize the real and imaginary part of the permittivity. Since ITO exhibits dielectric (transparent) behavior at shorter wavelengths than ENZ (~ 1250 nm) and metallic (lossy) behavior at longer wavelengths than ENZ (~ 1250 nm), a combination of Lorentz oscillator model and Drude model must be used to fit the ellipsometry data. Values of the amplitude ratio between in-plane, p, and out-of-plane, s, polarization, $\psi$, and fits at various angles of incidence are shown in Supplementary Figure 1. The noisy behavior around λ=1100 nm can be attributed to non-idealities of the sample and measurements that can be addressed using a B-spline fit that interpolates multiply connected polynomial functions and includes a bandwidth that models the smearing of wavelength at the detection spectrometer.

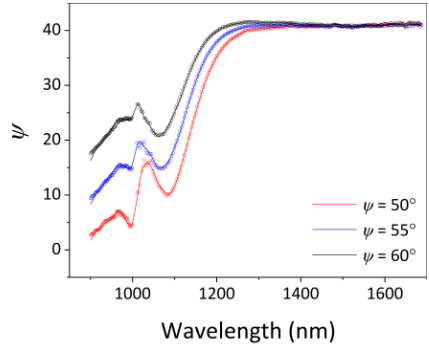

Supplementary Figure 1. Spectroscopic ellipsometry on ITO thin film. Experimentally measured values of $\psi$ using spectroscopic ellipsometry (circles) at three different angles of incidence. Fit with Lorentz oscillator model in the transparent region and Drude model in the lossy region. We employed B-spline fitting model that includes non- idealities of the experimental condition.

We fit the extracted dispersion from ellipsometry measurements by the Drude model described by

$$\varepsilon(\omega) = \varepsilon_\infty - \frac{\omega_p^2}{\omega^2 + i\gamma\omega}, \quad (1)$$

where $\omega_p = 2\pi \times 469 \times 10^{12}$ rad/s, $\gamma = 0.0410\omega_p$, and $\varepsilon_\infty = 3.65$ corresponds to a permittivity zero-crossing at $\lambda_{ENZ} \approx 1250$ nm. The plasma frequency can be tuned via controlling the doping rate [60] by inducing a non-parabolicity in the band structure and altering the plasma frequency through changes in the free-carrier density dependent effective mass of electrons, i.e., $\omega_p^2 = Ne^2/m_e^*\varepsilon_0$. The doping concentration calculated from the plasma frequency (extracted from ellipsometry) can be approximated to be $N \approx 7.63 \times 10^{26}\ m^{-3}$, while considering the effective mass of electrons for ITO to be $m_e^* = 0.28m_e$ [68]. The value of the free-electron density of ITO is around two orders of magnitude smaller than that of gold and silver [60, 68], resulting in a smaller plasma frequency compared to typical metals and consequently a dielectric to metal transition occurring at shorter wavelengths.

**Supplementary Note 2: Experimental techniques.**

We employed Beam-Deflection (BD) [40], cross-phase-modulation (XPM), and Z-Scan [36, 37] techniques for comprehensive characterization of the nonlinear optical properties of ITO thin-films. For our BD and XPM measurements, we used a regeneratively amplified Ti:Sapphire femtosecond laser system (775 nm) at a repetition rate of 1 kHz to pump an optical parametric amplifier (OPA) to generate near-infrared pulses (1000 nm - 1600 nm). For those measurements that the excitation pulse is centered at 775 nm, laser pulse is generated directly from our Clark-MXR CPA system. The pulsewidth of the pump at the output of the laser is ~ 150 fs (FWHM), and it is slightly chirped while propagating through various polarizing elements. The pulsewidth of the 775 nm pump beam at the sample plane is measured to be ~ 154 fs (FWHM). The probe beam was generated from an OPA system and tuned across the ENZ region. We kept the beam radius of probe at least × 4 smaller than that of excitation beam. For those measurements where the probe wavelength was between 1100 – 1550 nm, we directly used the signal beam, however, for the measurement performed at 950 nm, we used a BBO crystal to generate second harmonic of the idler beam at 1900 nm. For excitation in vicinity of ENZ, we use the same experimental setup and switch excitation and probe pulses and adjust the pulse energies accordingly. We measured the beam size of the near-infrared pulse using an InGaAs camera and fit the vertical and horizontal cross-sections with a Gaussian distribution. For each probe wavelength, we found the focus of the probe beam and placed the sample at that position. We also measured the excitation beam size and attempted to control the beam size to have a very close excitation beam size at each probe wavelength and subsequently to perform experiments under identical excitation conditions. For nearly degenerate experiments, where both excitation and probe pulses are near-infrared, we used a two-stage amplified Ti:Sapphire femtosecond laser system, also at 1 kHz repetition rate, (Coherent Legend Elite Due HE+) and pump two separate OPA systems to generate signal pulses.

In our experiments, the incident light first interacts with the ITO thin film and then propagates through the substrate. We chose this geometry due its simplicity in determining the incident light irradiance and subsequent analyses. We note that pulses experience multiple reflections within the ITO thin film, however, for the purpose of characterizing the NLO phase shift or NLR we can directly use the BD signal, $\Delta E/E$, and extract NLO parameters and only report the incident excitation irradiance.

## Supplementary Note 3: Beam-Deflection measurements for excitation in ENZ region and probe at $\lambda_p = 775$ nm.

This section includes BD measurements of ITO thin film for different near-infrared excitation wavelengths at various irradiances. The probe is kept at a much lower irradiance to only measure the NLO phase shift imposed upon probe via strong excitation pulse. We used the methodology explained in the main text to extract the BD signal of the substrate by directly subtracting total measured signal for parallel and perpendicular polarizations. In this method, we assumed that the dominant NLO mechanism contributing to ITO is due carrier redistribution and all instantaneous NLO effects are coming from the substrate. We performed similar measurements for oblique incidence ($\theta = 30$) to compare our results with normal incidence, as also explained in the main text. We also plotted the dependence of the peak $\Delta E/E$ versus excitation irradiance to verify the linear dependence of NLO phase shift on the irradiance, analogous to Kerr effect. We observed that the NLO response is saturated for $I_e > \sim 75$ $GW/cm^2$ at normal incidence.

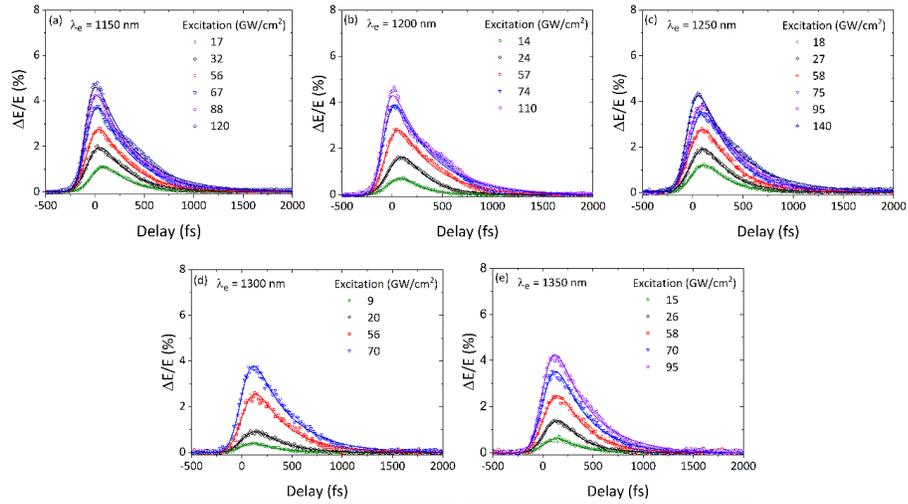

Supplementary Figure 2. BD measurements for a probe wavelength fixed at $\lambda_p = 775$ nm at normal incidence. (a-e) $\Delta E/E$ signals obtained for various excitation wavelengths covering the ENZ region at different excitation irradiances. Solid lines are fit to the experimental data obtained by considering exponential rise and fall time constants.

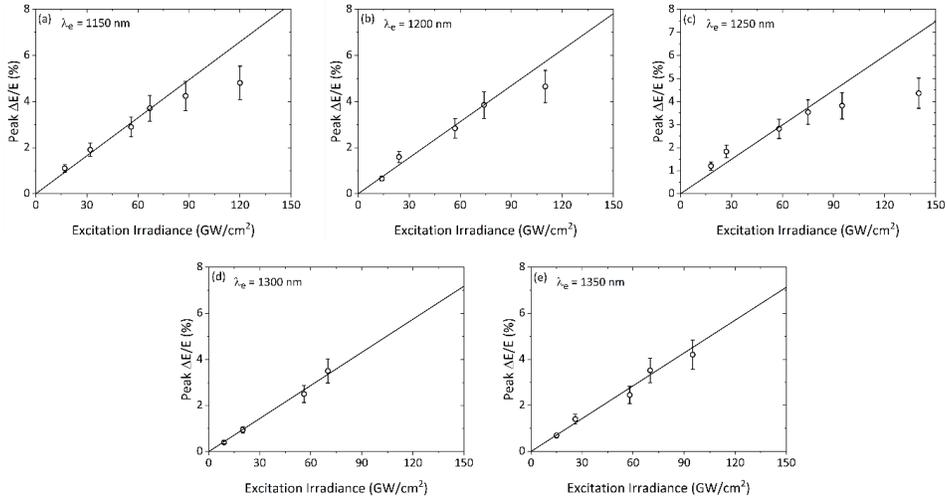

Supplementary Figure 3. Dependence of BD measurements on excitation irradiance at normal incidence. (a-e) Peak $\Delta E/E$ signals at $\tau_{D,max}$ versus excitation irradiance obtained for excitation wavelengths covering the ENZ region. The fitted solid line is directly related to the induced effective NLR.

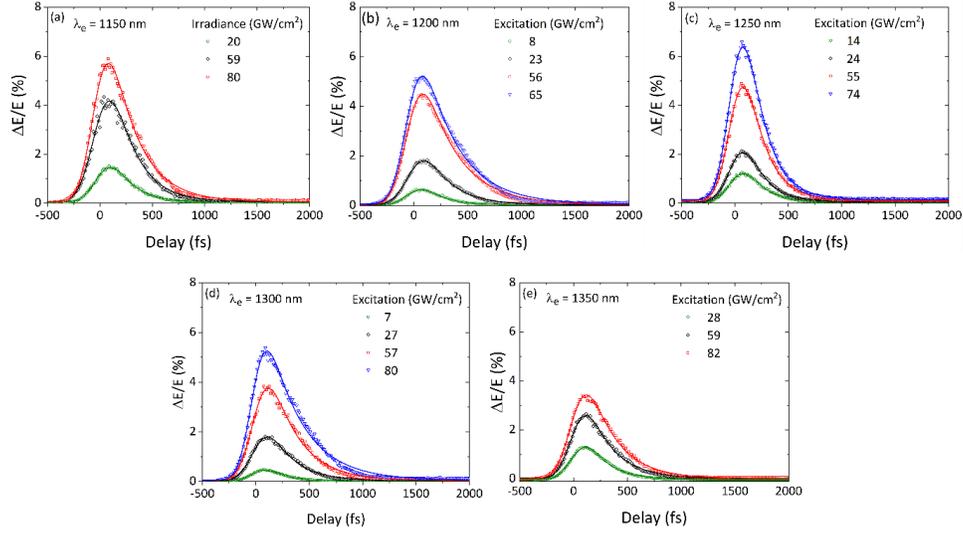

Supplementary Figure 4. BD measurements for a probe wavelength fixed at $\lambda_p = 775$ nm at oblique incidence ($\theta = 30$ deg). (a-e) $\Delta E/E$ signals obtained for various excitation wavelengths covering the ENZ region at different excitation irradiances. Solid lines are fit to the experimental data obtained by considering exponential rise and fall time constants.

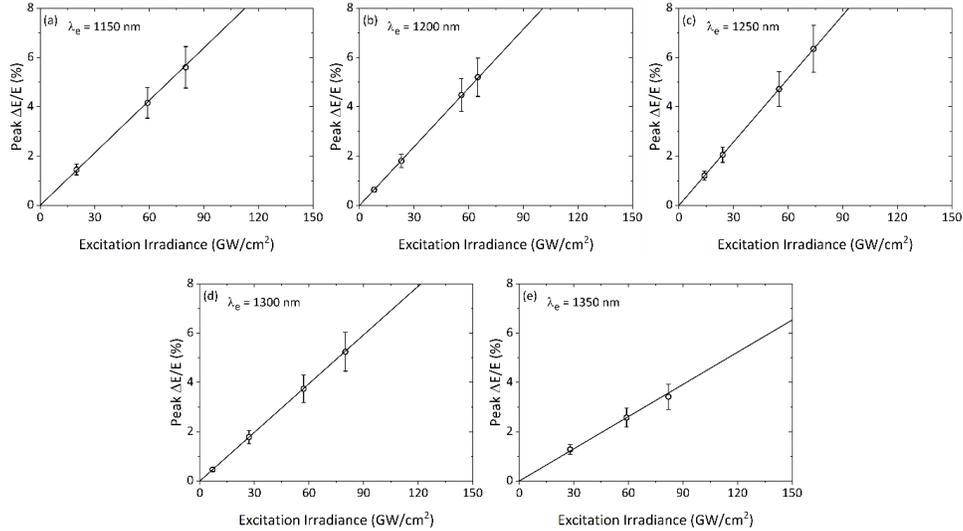

Supplementary Figure 5. Dependence of BD measurements on excitation irradiance at oblique incidence ($\theta = 30\ deg$). (a-e) Peak $\Delta E/E$ signals at $\tau_{D,max}$ versus excitation irradiance obtained for excitation wavelengths covering the ENZ region. The fitted solid line is directly related to the induced effective NLR.

## Supplementary Note 4: Beam-Deflection and XPM measurements for probe in ENZ region and excitation at $\lambda_e = 775\ nm$.

This section includes BD measurements of ITO thin film for different near-infrared probe wavelengths for an excitation beam at 775 nm. We used similar analysis approach to previous section and extracted the BD signals for ITO thin films at normal incidence. However, in these measurements we also measured XPM at different delays and compared the induced shift in probe wavelength with predictions calculated from our BD measurements. Similarly, we also plotted the dependence of the peak $ΔE/E$ versus excitation irradiance to verify the linear dependence of NLO phase shift on the irradiance.

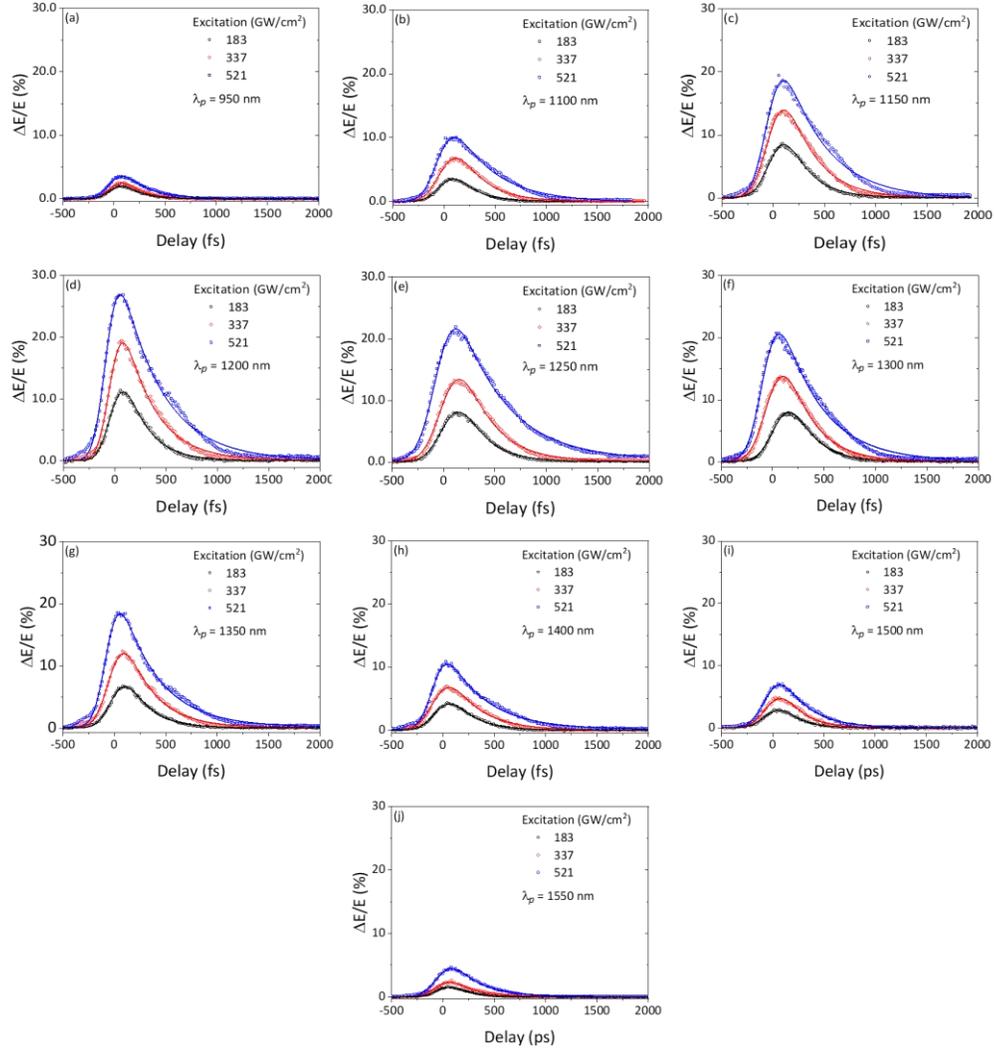

Supplementary Figure 6. BD measurements for excitation at 775 nm and probe at $λ_p$. (a-j) $ΔE/E$ (%) versus delay of different probe wavelengths (listed in the figure) directly measured by BD and after eliminating the effect of the substrate while excited at $λ_e = 775$ nm, $I_e = 183\ GW/cm^2$ (black), $I_e = 337\ GW/cm^2$ (red) and $I_e = 521\ GW/cm^2$ (blue). Solid lines are fit to the experimental data obtained by considering exponential rise and fall time constants.

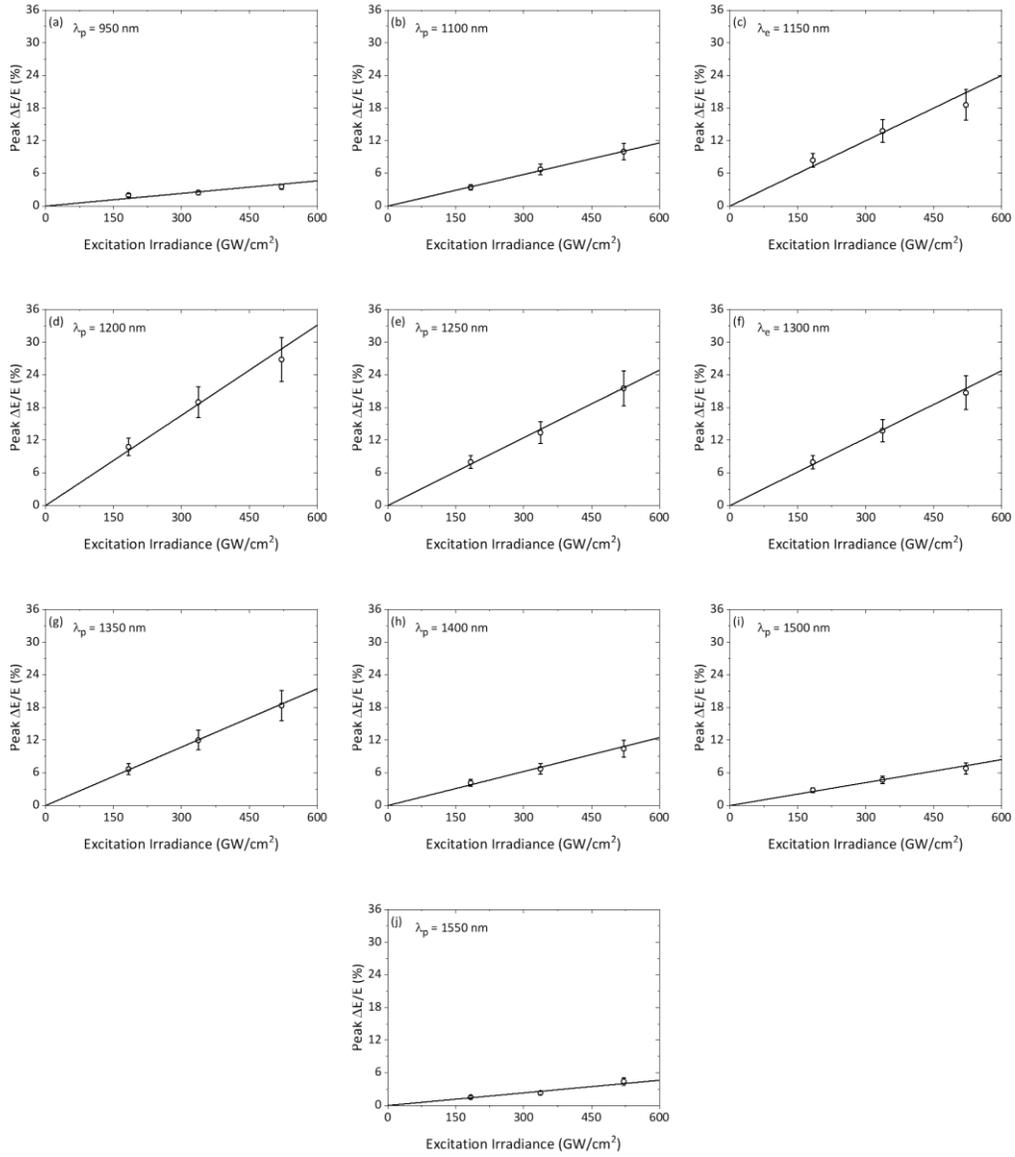

Supplementary Figure 7. Dependence of BD measurements for excitation at 775 nm and probe at $\lambda_p$ (labeled in each plot) on excitation irradiance at normal incidence. (a-j) Peak $\Delta E/E$ signals at $\tau_{D,max}$ versus excitation irradiance obtained for probe wavelengths covering the ENZ region. The fitted solid line is directly related to the induced effective NLR.

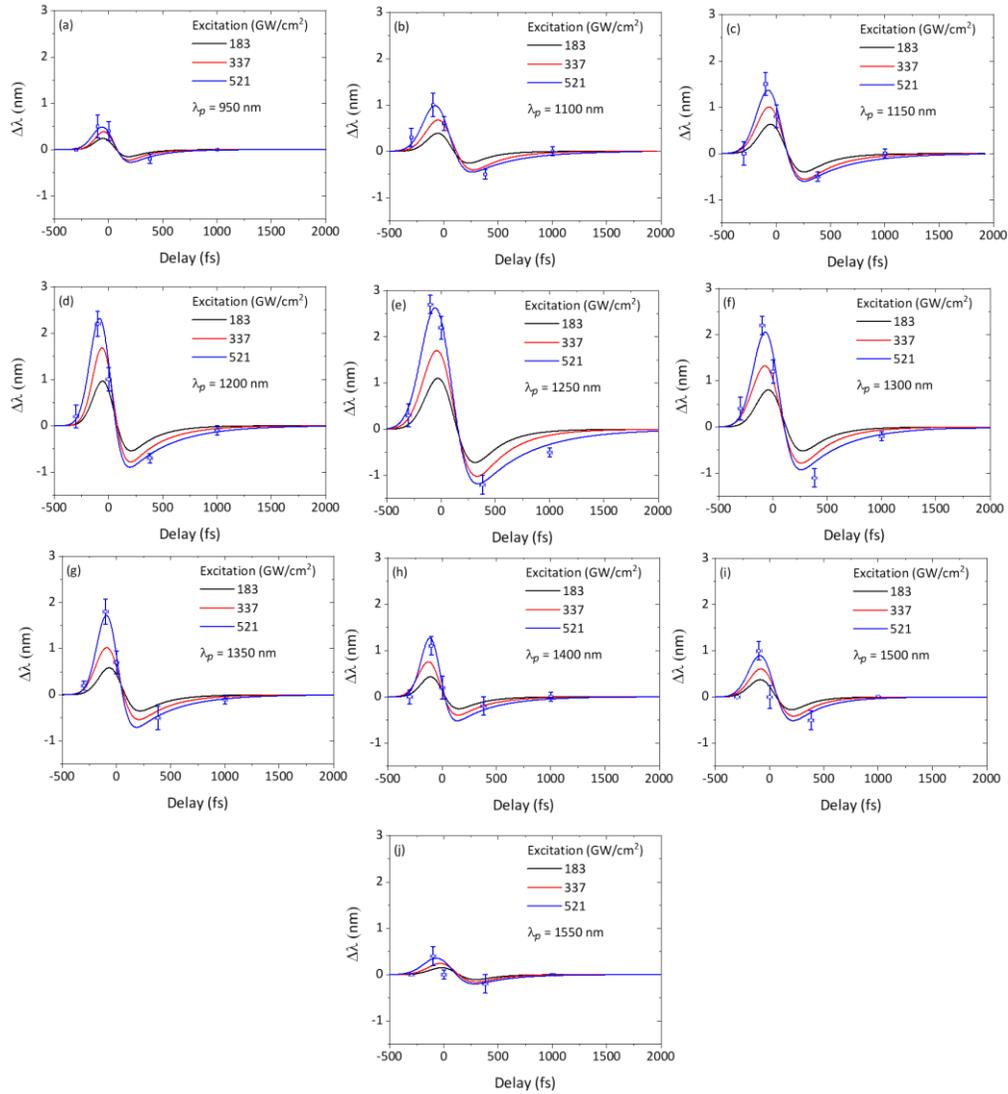

Supplementary Figure 8. Optically induced spectral shift of the probe pulse induced by excitation at 775 nm. (a-j) Magnitude (solid lines) of the spectral shift imposed upon probe at $\lambda_p$ (labeled in each figure) versus relative delay between excitation and probe pulse calculated directly from BD measurements for various excitation irradiances. Dataset at each probe wavelength is directly measured via cross-phase-modulation experiment by using an optical spectrum analyzer.

## Supplementary Note 5: Beam-Deflection and XPM measurements for probe in ENZ region and excitation at $\lambda_e = 1250$ nm.

In this section we present BD and XPM measurements for near-infrared excitation and probe pulses. In these experiments, we fixed excitation wavelength at $\lambda_e = 1250$ nm generated from a high energy OPA system (TOPAS-HE) and tune the probe pulse across ENZ region. The probe pulse is generated from another OPA system (TOPAS-Prime) and attenuated accordingly.

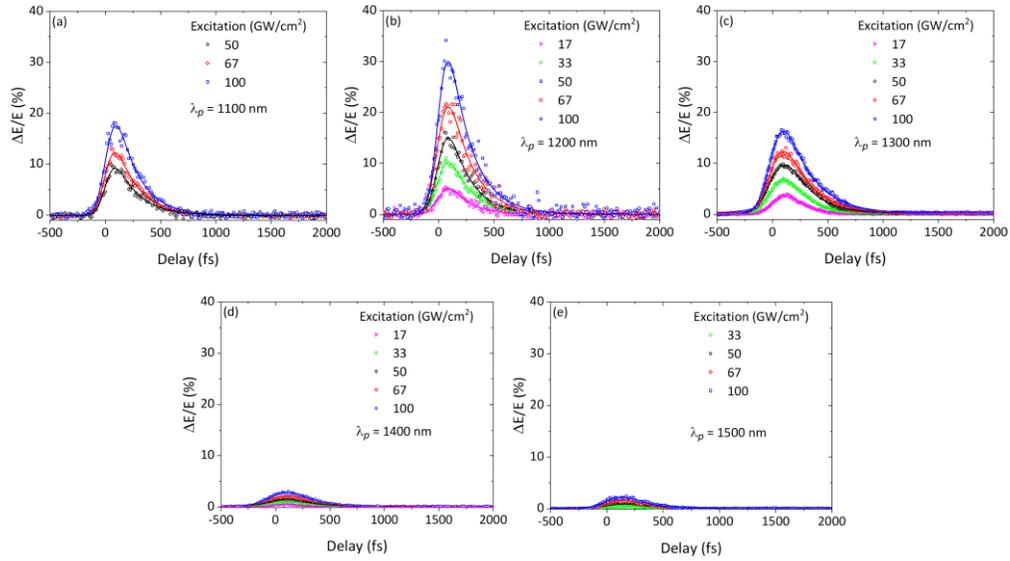

Supplementary Figure 9. BD measurements for excitation at 1250 nm and probe at $\lambda_p$. (a-e) $\Delta E/E$ (%) versus delay of different probe wavelengths (listed in the figure) directly measured by BD and after eliminating the effect of the substrate while excited at $\lambda_e = 1250$ nm, $I_e = 17\ GW/cm^2$ (magenta), $I_e = 33\ GW/cm^2$ (green), $I_e = 50\ GW/cm^2$ (black), $I_e = 67\ GW/cm^2$ (red), and $I_e = 100\ GW/cm^2$ (blue). Solid lines are fit to the experimental data obtained by considering exponential rise and fall time constants.

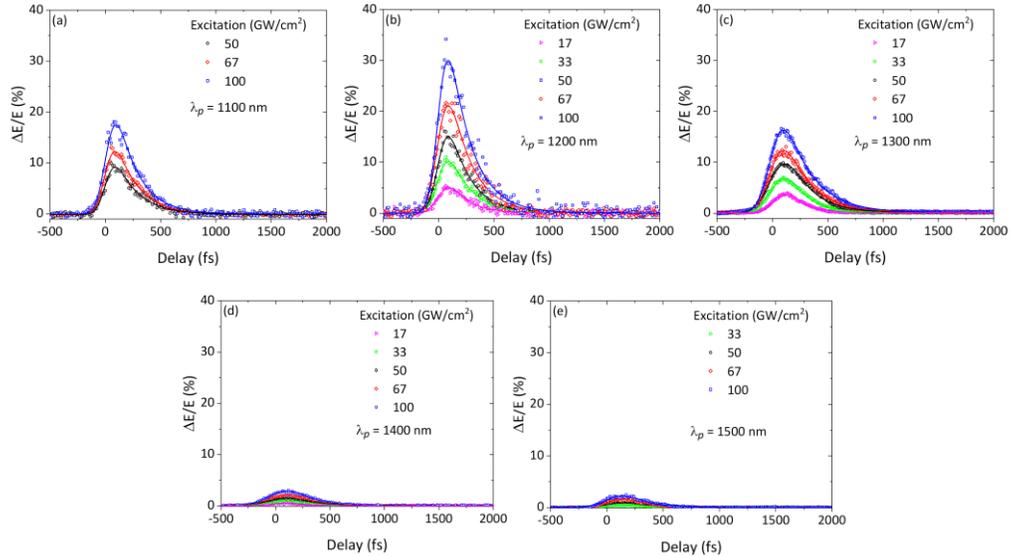

Supplementary Figure 10. Dependence of BD measurements for excitation at 1250 nm and probe at $\lambda_p$ (labeled in each plot) on excitation irradiance at normal incidence. (a-e) Peak $\Delta E/E$ signals at $\tau_{D,max}$ versus excitation irradiance obtained for probe wavelengths covering the ENZ region. The fitted solid line is directly related to the induced effective NLR.

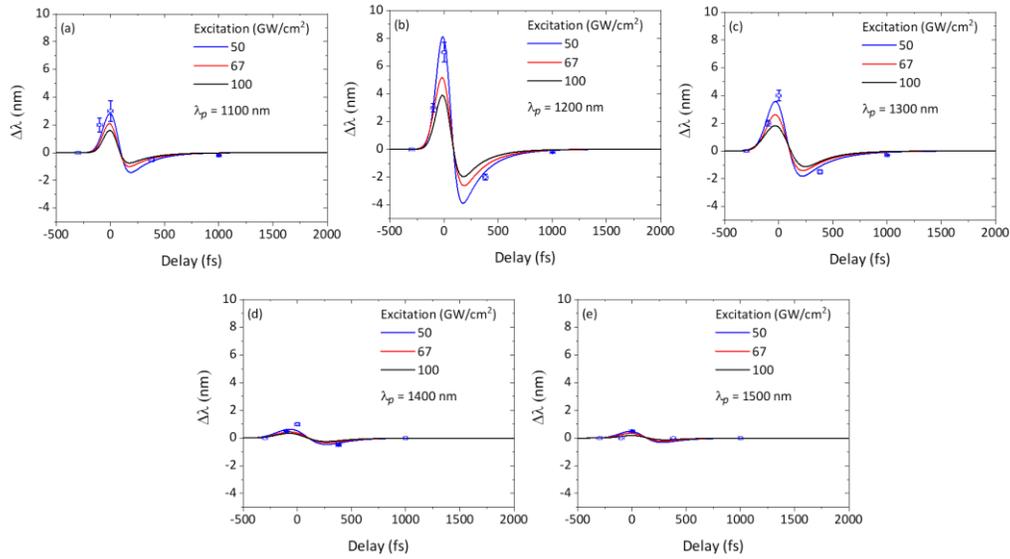

Supplementary Figure 11. Optically induced spectral shift of the probe pulse induced by excitation at 1250 nm. (a-e) Magnitude (solid lines) of the spectral shift imposed upon probe at $\lambda_p$ (labeled in each figure) versus relative delay between excitation and probe pulse calculated directly from BD measurements for various excitation irradiances. Dataset at each probe wavelength is directly measured via cross-phase-modulation experiment by using an optical spectrum analyzer.

## Supplementary Note 6: Z-Scan measurements to characterize degenerate NLR of ITO thin film.

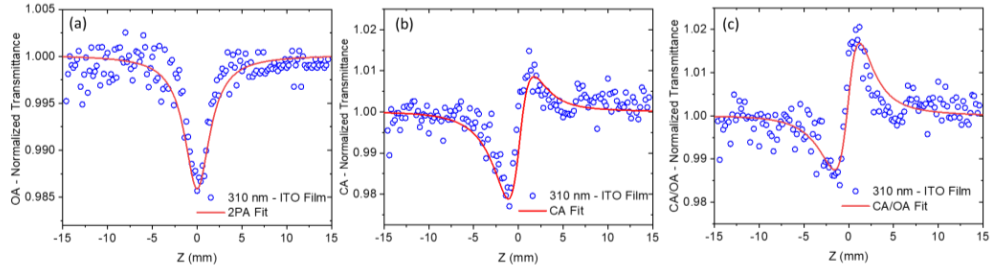

Supplementary Figure 12. Degenerate Z-Scan measurements at $\lambda = 775$ nm. Measurements performed at $\lambda = 775$ nm, at $I_0 = 368\ GW/cm^2$, (a) open-aperture (OA) Z-Scan, (b) closed-aperture (CA) Z-Scan, and (c) CA/OA. Two-photon absorption and nonlinear refraction fit with values of $\alpha_2 = (3.8 \pm 0.5)\ cm/GW$ and $n_2 = (1.3 \pm 0.2) \times 10^{-4}\ cm/GW^2$.

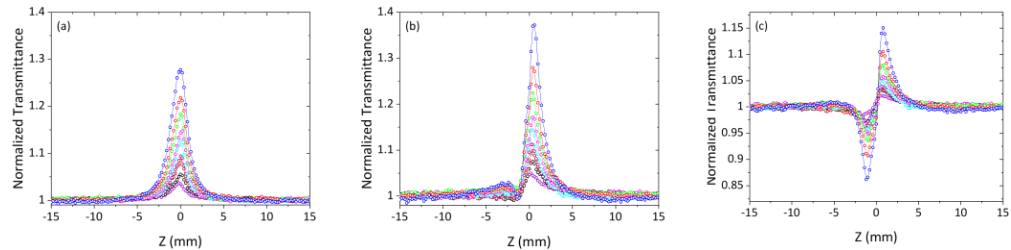

Supplementary Figure 13.s Degenerate Z-Scan measurements at $\lambda = 1250$ nm. Measurements performed at $\lambda = 1250$ nm, for various irradiances ($I_0 = 10, 16, 24, 31, 40, 46, 58, 78, 113\ GW/cm^2$), (a) open-aperture (OA) Z-Scan, (b) closed-aperture (CA) Z-Scan. The NLR is deduced directly from $\Delta T_{pv}$ obtained from (c) CA/OA [36, 37].